# A New Type "Magnetic-field" Probe of High Spatial-resolution Based on a Single-layer Flat-coil Method

S.G. Gevorgyan[1,2] and M.G. Azaryan[3]

[1]*Department of Solid State Physics, Yerevan State University, 1 Alex Manoogian str., 375025, Yerevan, Armenia*
[2]*Institute for Physical Research, National Academy of Sciences, Gitavan IFI, 378410, Ashtarak-2, Armenia*
[3]*Department of Semiconductor Physics & Microelectronics, Yerevan State University, 1 Alex Manoogian str., Yerevan, Armenia*

*Abstract*
A radically new approach to surface probing based on a replacement of the solid-state near-field probes by the "long-field" ones is presented and discussed. Such probes may enable to create a radically new generation microscopes with non-perturbing long-range action probe based on a highly sensitive RF test method with a single-layer flat coil. This may give a start for creation of a new direction in microscopy.

**1. Introduction**

Since the invention of the first scanning **tunneling microscope** (**TM**) by Heinrich Rohrer and Gerd Binning in 1981 [1-2], scanning probe microscopy has enabled a burst of nanotechnology achievements that includes the manipulation and arrangement of individual atoms on a surface.

Scanning ***probe microscopes*** (**PM**) allow scientists to visualize, characterize and even manipulate material structures at extremely small scales, including features of atomic proportions. A wide variety of material structures and properties can be studied in man-made and natural systems, including biological systems. The family of scanning **PM** uses no lenses, but a probe that interacts with the sample surface. The type of interaction measured between the probe tip and the sample surface determines the type of scanning probe microscope being used.

Strong dependence of a detected signal on the size of spatial-gap ($Z$) between the solid-state probe and surface of the object under test is the main physical principle of a probe microscopy [3]. That lies on the base of operation both tunneling [1-2, 4] and ***atomic-force microscopes*** (**AFM**) [5]. In tunneling microscopes the measured signal (*tunneling current*) drops exponentially with a gap ($I(Z)\sim\exp(-kZ)$), while, in **AFM** the signal, $S$, taken from the probe-position sensor of the *cantilever*, drops far stronger with a gap increase ($S(Z)\sim Z^{-6\div7}$). The probes of microscopes are solid-state needles with tips of a nanometer-size radius of curvature. In **TM**-microscopes the probe is electrochemically pointed wire usually, made from the tungsten, platinum-iridium or platinum-rhodium, while in **AFM** – *cantilever* (fixed from the end elastic sensitive arm, having high value of the resonance frequency of its mechanical oscillations, with a pyramid-form tip on its next free end of a necessary radius of curvature, created by the microelectronic processing).

Interest to **TM** and **AFM** microscopes is caused first of all by their unique resolution – up to a few hundredth parts of angstrom along normal to the surface of object under test, and a few units of angstrom – across it. At that, there is no need of high vacuum at all for the operation of such a microscope (as that is the case in all electron microscopes) – they may operate both in air and in a liquid environment.

Working gap between detecting probe and the object is small in acting probe-microscopes ($Z$<1nm). Due to this one has a lot of difficulties. In particular, it is impossible practically to avoid collision between the probe and surface of the object under test [1-6]. This usually leads also to a damage of the needle's tip, and thus, results in a worsening of its resolution, and ultimately, to distortion of the obtained information.

***Imperfections of the probe-microscopes are as follows:***

**a)** *uncontrolled thermo-extension of a probe* (especially, at the tunneling modes of operation of the microscopes – because of its ohmic heating). Due to this an image of the object surface becomes distorted. Its correction needs special technical and software complication of the microscopes [7];

**b)** *perturbations inserted into the object under test,* caused by the tunneling currents (for example, destruction of Cooper pairs in superconductive material under test – especially, near the phase transition).

In this connection note that due to organization of an interaction with the object by a slowly disturbing electromagnetic field another type of the probe-microscope – the so-called **near-field** (**NF**) microscope – is practically non-disturbing. And, although in case of the optical NF (**ONF**) microscopes the field comes out from the pointed end of the fiber-optic probe [8], covered by the metallic layer, but the resolution even in such way improved microscopes may not reach nano-scales, in a principle;

**c)** *small work-distances between the probe and surface of the object* (~1nm), which in many respects limits technically the abilities of explorer to apply various types of testing perturbations, if needed.

**d)** *solid-state probe*-containing *devices need special manufacture* (cantilevers, probes of the NF-microscopes, etc. [9-10]). Besides, there is need to get also a large number of such accessories, for their effective replacement during the exploitation period of the microscopes – due to their very often breakage.

**e)** *difficulties related with the registration of informative signal* (especially in **ACM**-microscopes).

To overcome above difficulties in a probe-microscopy one needs the formation of radically new conceptions in this area, as well as creation of the laboratory models for their realization. At a best, the energy of the scholars should be directed (in our opinion) to reveal the physical principles promoting creation of **next-generation probes** – *"far-ranging", "non-disturbing", non-contact, no special manufacture, and user-friendly…* Presented here technique is worthy of attention not only in respect with the listed advantages, but also for its future perspectives – to be used also for other applications too. Keeping all the advantages of acting presently *near-field* probe-microscopes [1-5, 8-10 ] we are offering, at the same time, the possibility to replace solid-state probes by the ***non-solid and "far-ranging"*** ones, with the regulated metrological characteristics. Perhaps, *this may put the start of a new-generation microscopy*.

**2. Preliminary test-results and their discussion**

*We developed a highly sensitive test-method* for detecting small changes in magnetic inductance of the object under the test (**~$10^{-12}$H**). It is suitable also for detecting the small amounts of the power, released in or observed by the specimen (**~$10^{-9}$W**). The method is *based on* a low-power tunnel diode (**TD**) oscillator with *a single-layer open-flat geometry coil* [11-12] (see Fig.1).

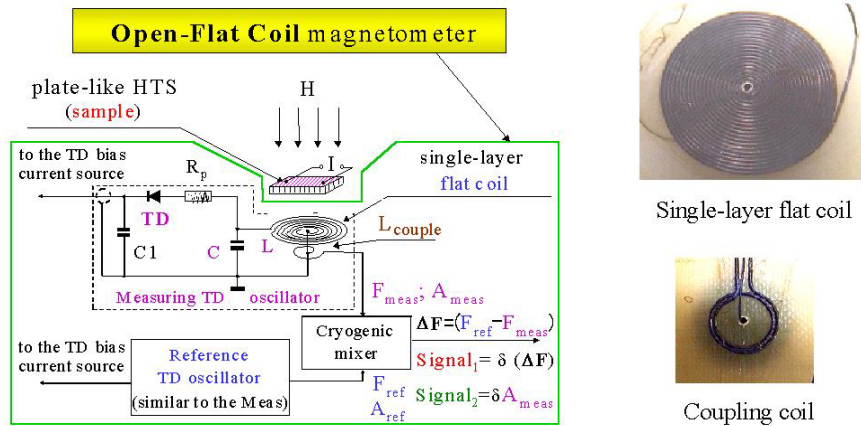

**Fig.1.** Schematics of the method based on low-power tunnel diode oscillator with a single-layer open-flat geometry coil [11-12].
*Side insets:* single-layer flat geometry pick-up coil ($\Phi_{coil}=2R_{coil}$ ~14mm), and the coupling coil ($\Phi_{couple}$~5mm).

As is seen from experimental data below, the *flat coil-based technique enables* above crucial strong dependence of detected signals from the *Z*, which gives *a real possibility for creation of practically "non-disturbing"* (<1μW) *"magnetic-field"* (**MF**) *probes of a radically new type*. Note that the gap *Z* between such *a probe-formative* **coil** and the object under test may be by orders of magnitude larger, compared to all known microscopes. ***Such* MF**-probe *may be used* in future (after proper adaptation of a design of acting microscopes to the new probe) *for creation of a radically new principle of operation microscopes*, capable of probing and analyzing atomic-scale objects, *with small* (*but*/and *adjustable*) *perturbations* for the object under test. Such a microscope may enable also to distinguish magnetic atoms from non-magnetic ones, which is important, for example, to study the coexistence of superconductivity and magnetism – detected, at the moment, in *sub*-micron scale objects only.

*The* **MF**-probe's *preliminary test-results are presented in Figs. 2-3.* Note that, the frequency, *F*, and amplitude, *A*, of TD oscillator are used in such a technique as testing parameters. At that, the measuring effects are determined by a distortion of testing MHz-range field's configuration near pick-up coil [11-12], and by absorption of the same field's power by a sample under test [13-14] (due to various external influences). This finally results in the changes of TD oscillator frequency or/and amplitude respectively.

Figure 2 shows the measured dependences of the frequency *F* and amplitude *A* of the oscillator on lateral position of ~1mm in diameter metallic tin (**Sn**) test-ball relative to the face of the *probe-formative* **coil** (**MF**-probe). At movement the ball distorts RF-testing field of the coil, causing itself also absorption of the same field's power, thereby leading to the changes of the oscillator frequency and amplitude. The tests are made on the mechanical ***XYZ***-translation stage described in [15]. It provides manual micrometer-positioning for the test ball relative to the surface of the probe-formative flat coil ($\Phi_{coil}=2R_{coil}$ ~14mm, and the gap between the ball and the flat coil was $Z_0$=1.5mm).

The shapes of the presented curves (*they look like to the probe*) along the diameter of the coil (per se, these are the RF-testing field lines of the coil) obey the laws close to $F\sim1/(R_{coil})^2$, $A\sim1/(R_{coil})^2$, which well agree with the coil calibration data presented in [12]. According to data presented in Fig.2 one may estimate lateral spatial sensitivity for our **MF**-probe of about $\gamma_F$ ~0.1μm/(±3Hz) and $\gamma_A$ ~0.4μm/(±10μV). Taking into account the frequency and amplitude metering errors in our tests (±3Hz and ±10μV respectively) one may conclude: even for such large sizes of the probe-formative coil ($\Phi_{coil}$ ~14mm) we could reach the lateral spatial resolution of about $\delta_F$ ~100nm by frequency, and $\delta_A$ ~ 400nm by amplitude tests.

The next Figure 3 demonstrates measured dependences for the frequency $F$ and amplitude $A$ of the oscillator on the size of gap $Z$ between the probe-formative coil and test-ball ($Z$ was changed from 1.5mm to 3.0mm during our tests, with a step of about ~100μm). The analysis of the curves' shapes shows that $Z$ dependences of $F$ and $A$ are exponential – $F(Z)\sim\exp(-k_F Z)$ and $A(Z)\sim\exp(-k_A Z)$, which also agrees with our coil calibration data presented in [12]. According to these data one may estimate the spatial-sensitivity for our **MF**-probe along its $Z$ axis of about $\gamma_F(Z)$~0.03μm/(±3 Hz). Taking into account the mentioned frequency error of about ±3Hz, one may estimate the spatial-resolution of **MF**-probe $\delta_F$~30nm, reached at a distance of ~1-2mm from the face of the coil, along its $Z$ axis – detected by frequency measurements.

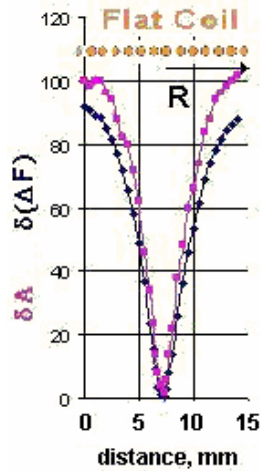 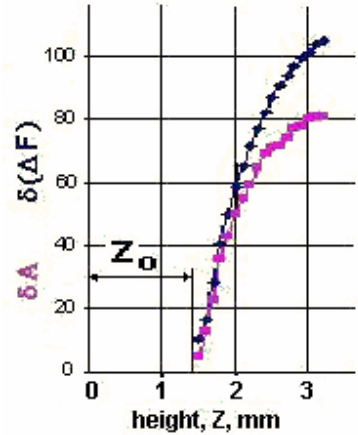

**Fig.2.** The measured dependences of the frequency $F$ and amplitude $A$ of the tunnel diode oscillator on the lateral position of ~1mm in diameter metallic tin (**Sn**) test-ball relative to the face of a single-layer flat coil (**MF**-probe).

**Fig.3.** The measured dependences of the frequency $F$ and amplitude $A$ of the tunnel diode oscillator on the gap size $Z$ between the **MF**-probe (flat coil) and the test-ball. ($Z$ was changed from 1.5mm to 3.0mm during our tests, with a step of about ~100μm).

There are a few real possibilities *how to improve spatial resolution* (including, lateral) *of such a new* far-ranging "magnetic-field" *probe* – to reach the *sub*-nanometer scales. One of the ways is to apply the modulation for the position of the object under tests (*or/and* for the size of the gap) with a subsequent use of lock-in amplification technique's advantages (for both, the frequency and amplitude signals, independently). Another possibility relates with a *decrease of the probe-formative coil's geometrical sizes* – up to at lease a few hundreds of micrometers, in diameter. The combination of these two ways may undoubtedly bring much better results.

*Expected advantages of the microscopes with a new* "*magnetic-field*" (MF) *probes are as follows:*
**a)** *absence of the uncontrolled thermo-extension of a probe*, peculiar to **TM**-microscopes;
**b)** practically complete *absence of the perturbations inserted into the object by the probe*. *The possibility of inserting*, if needed, *of the small regulated perturbation*;
**c)** *incomparably large work-distances (≥10μm) between the probe and surface of the object*, which may enable visually to control the local area of the investigation in object under the test. And, if needed, to be able to apply various types of test-perturbations (all acting probe-microscopes need, in principle, the work-distances of about a few angstroms to nanometer scales).
**d)** *the nature of interaction of a new* **MF**-probe *with the objects under test enables,* in principle*, the possibility of manipulation by nano-scale objects on the object surface* – to be confirmed empirically;

**e)** *simplicity of the technique related with a registration of the informative signal* (and, incomparably high relative resolution (better $10^{-6}$), provided solely by the flat coil-based highly sensitive method).

In conclusion, we consider necessary to note also, that in order to be able to curry out all the above mentioned scientific-technical and technological investigations, as well as taking account our wide experience [7, 11-14, 16-19], we give an important role to the use of *LabVIEW* – with related hardware-based solutions of National Instruments – as an optimally developed and adapted programming medium for effective conducting of the modern research work.

**REFERENCES**
1. Binning G. and Rohrer H., Scanning Tunneling Microscope, *US Patent 4,343,993 Aug.10*, *1982*. Field: Sep.12, 1980.
2. Binning G. et al., Appl.Phys. Lett. 40, 178 (1982).
3. Bikov V.A., Russian *doctoral degree dissertation*, Moscow (2000) – *topical review* on probe microscopes [in Russian].
4. Binning G., Rohrer H., *UFN*, **154** (1988) 261 [in Russian].
5. Binning R., Quate C.F., Gerber Ch., *Phys. Rev. Lett.*, **56** (1986) 930.
6. Azaryan M.H., Proc. of 3-rd National Conference, Sevan, Armenia, pp.283-287 (2001) [in Russian].
7. Azaryan M.H., Haroutyunyan V.M., et al., Proc. of the Conf. on "Educational, Scientific & Technological Appl. in *LabVIEW Environments*, and *Technologies of National Instrum.*", Moscow (2004).
8. Dryakhlushin V.F., Klimov A.Yu. et al., *PTE* , **2** (1998) 138-139 (Russian *Rev. Sci. Instrum.*).
9. Edelman V.S., *PTE* , **1** (1991) 24-42 (Russian *Rev. Sci. Instrum.*) - review.
10. Volodin A.P., "News in Scanning Microscopy", *PTE* , **6** (1998) 3-42 (Russian *Rev. Sci. Instrum.*) – review on the materials of the International Conference (**STM'97**).
11. Gevorgyan S.G., Kiss T., et al., *Rev Sci. Instrum.*, **71**(3) (2000) 1488-1494.
12. Gevorgyan S.G., et al., *Physica C:* "Superconductivity & Applications", **366**(1) (2001) 6-12.
13. Gevorgyan S.G., et al., *Physica C:* "Superconductivity & Applications", **363** (2001) 113.
14. Gevorgyan S.G., Kiss T., et al., *IEEE Trans. on Appl. Supercond.*, **11** (2001) 3931.
15. Azaryan M., Adamyan Z., Proc. of 2-nd Nat.l Conf., Dilizhan, Armenia, 237-240 (1999) [in Russian].
16. Gevorgyan S.G., Kiss T., et al., *Supercond. Sci. Technol.*, **14** (2001) 1009-1014.
17. Gevorgyan S.G., et al., *Physica C:* "Superconductivity & Applications", **378-381** (2002) 531.
18. Gevorgyan S.G., Kiss T., et al., *Supercond. Sci. Technol.*, **14** (2001) 1009-1014.
19. Gevorgyan S.G., et al., *Physica C:* "Supercond. & Appll.", **378-381** (2002) 531.